\documentclass[twocolumn,pra,superscriptaddress]{revtex4}
\pdfoutput=1
\usepackage{amssymb}
\usepackage{amsmath}
\usepackage{graphicx,bm}
\usepackage{epstopdf}
\usepackage{subfigure}
\usepackage{natbib}
\usepackage{epsfig}
\usepackage{amsfonts}
\usepackage{mathrsfs}
\usepackage{CJK}
\usepackage[toc,page,title,titletoc,header]{appendix}
\usepackage{array,longtable}

\usepackage[usenames, dvipsnames, x11names]{xcolor}

\usepackage[colorlinks]{hyperref}
\hypersetup{
 colorlinks=true,
 citecolor=red,
 linkcolor=blue,
 urlcolor=cyan}

\newcommand{\hxi}{\hat{\xi}}

\begin{document}
\title{Spectral characterization of non-Gaussian quantum noise: Keldysh approach and application to photon shot noise}

\author{Yu-Xin Wang}
\affiliation{Pritzker School of Molecular Engineering,  University  of  Chicago, 5640  South  Ellis  Avenue,  Chicago,  Illinois  60637,  U.S.A.}

\author{A. A. Clerk}
\affiliation{Pritzker School of Molecular Engineering,  University  of  Chicago, 5640  South  Ellis  Avenue,  Chicago,  Illinois  60637,  U.S.A.}

\date{\today}

\begin{abstract}
Having accurate tools to describe non-classical, non-Gaussian environmental fluctuations is crucial for designing effective quantum control protocols and understanding the physics of underlying quantum dissipative environments. We show how the Keldysh approach to quantum noise characterization can be usefully employed to characterize frequency-dependent noise, focusing on the quantum bispectrum (i.e., frequency-resolved third cumulant).  Using the paradigmatic example of photon shot noise fluctuations in a driven bosonic mode, we show that the quantum bispectrum can be a powerful tool for revealing distinctive non-classical noise properties, including an effective breaking of detailed balance by quantum fluctuations.  The Keldysh-ordered quantum bispectrum can be directly accessed using existing noise spectroscopy protocols.     
\end{abstract}
%\pacs {}

\maketitle

%%%%%%%%%%%%%%%%%%%%%%%%%%%%%%%%%%%%%%%%%%%%%%%%%%%
\section{Introduction}
An accurate description of environmental fluctuations is crucial for quantum information processing and quantum control. While it is common to assume noise that is both classical and Gaussian, there are many physically relevant situations where these assumptions fail~\cite{Cywinski2014,Viola2014,Viola2016,Oliver2019,Ramon2019}. 
%(e.g.~$1/f$ noise from a bath of two-level fluctuators, the effective dephasing from hyperfine coupling to an ensemble of nuclear spins).  
Understanding how to usefully characterize non-Gaussian, non-classical noise in a frequency-resolved manner could enable the design of more optimal dynamical decoupling protocols, enhancing qubit coherence.  It could also provide fundamental insights into the nature of the underlying dissipative environment.

For classical noise, the frequency-resolved higher noise cumulants (so-called polyspectra \cite{Rao1984book}) provide a full characterization.  These have been previously measured for classical non-Gaussian fluctuations in a single-electron transistor \cite{Haug2012}.  
More recent work
has proposed \cite{Viola2016,Ramon2019} and demonstrated \cite{Oliver2019} 
protocols to reconstruct polyspectra using a qubit driven by classical non-Gaussian noise; 
Ref.~\cite{Viola2016} also studied the specific class of linearly coupled oscillator baths, where operator ordering does not play a role~\cite{footnote1}. The full generalization of these ideas to quantum non-Gaussian noise as produced by a 
{\it generic} quantum environment
(i.e., one where operator ordering matters)
remains an interesting open question; for this general problem, the non-commutativity of noise operators at different times poses a challenge as to how one should appropriately define polyspectra.

In this paper, we show that the Keldysh approach \cite{Levitov1996,Nazarov2003,BelzigPRL2010,Clerk2011,Hofer2017}, a method used extensively to characterize low-frequency noise, also provides an unambiguous and practically useful way to describe non-Gaussian quantum bath noise in the frequency domain.
It provides a systematic way to construct a quasiprobability distribution to describe the noise, and to assess whether the noise can be faithfully mimicked by completely classical noise processes \cite{Nazarov2003,Clerk2011}. 
It also has a direct operational meaning:  the ``quantum polyspectra" we introduce are {\it exactly} the quantities that contribute to the dephasing of a coupled qubit at each order in the coupling.  Moreover, these quantities can be measured using {\it the same} non-Gaussian noise spectroscopy techniques designed for classical noise sources \cite{Viola2016,Oliver2019,RBLiu2019}; one does not have to decide in advance whether the noise is classical or quantum to perform the characterization.  
Note that a recent work presented a method to measure arbitrary quantum bath correlation functions \cite{RBLiu2019}; in contrast, our work focuses on characterizing the most physically relevant correlation function at each order and identifying a corresponding quasiprobability.

To highlight the utility of our approach, we apply it to the concrete but non-trivial case of photon shot noise in a driven-damped bosonic mode (a relevant source of dephasing noise in circuit QED systems~\cite{Schoelkopf2006,Devoret2019} among others).  Prior work used the Keldysh approach to study this noise at zero frequency \cite{Clerk2011,Clerk2016}; here we instead focus on the behaviour of the frequency-resolved third cumulant, the ``quantum bispectrum" (QBS).  We show that the QBS reveals important new physics and distinct quantum signatures:  at low temperatures, qualitatively new features emerge that would never be present in a classical model with only thermal fluctuations.  We also show that the QBS is a generic tool for revealing the breaking of detailed balance and violation of Onsager-like symmetry relations.  We find that the photon shot noise QBS violates detailed balance at low temperatures.

%%%%%%%%%%%%%%%%%%%%%%%%%%%%%%%%%%%%%%%%%%
%%%%%%%%%%%%%%%%%%%%%%%%%%%%%%%%%%%%%%%%%%

\section{Keldysh ordering and quantum polyspectra}

Consider first a classical noise process $\xi(t)$.  Its moment generating function (MGF) is defined as
\begin{align}
\label{Eq:MGFcl}
    \Lambda_{\rm class}[F(t); t_f] & = \overline{ \exp \left[ -i \int_0^{t_f} F(t) \xi(t) \right] },
\end{align}
where the bar indicates a stochastic average.  Functional derivatives of 
$\Lambda_{\rm class}$ with respect to $F(t)$ can be used to calculate arbitrary-order correlation functions of $\xi(t)$, while functional derivatives of $\ln \Lambda_{\rm class}$ generate the cumulants of $\xi(t)$ (see, e.g., Ref.~\cite{JacobsStochasticBook}).  Fourier transforming these cumulants yields the polyspectra, which completely characterize the noise in the frequency domain \cite{Rao1984book}.  
% Up to the usual caveats, this completely characterizes the probability distribution of the noise.

In the quantum case, our noise is a Heisenberg picture operator $\hat{\xi}(t)$ whose evolution is generated by the Hamiltonian of some bath; we take $\hat{\xi}(t)$ to be Hermitian for simplicity.  Defining correlation functions now has some subtlety, as $\hat{\xi}(t)$ will not in general commute with itself at different times; hence, different time-ordering choices yield different results.  Correlation functions at a given order describe both how the bath responds to external perturbations, as well as its intrinsic fluctuations \cite{Kamenev2011book}.  We are interested here in characterizing the latter quantity, and asking whether these fluctuations are equivalent to an effective classical noise process.

The well-developed machinery of Keldysh quantum field theory provides a precise method for accomplishing our task \cite{Levitov1996,Nazarov2003,BelzigPRL2010,Clerk2011,Hofer2017}.  While this approach is completely general, the simplest derivation is to imagine coupling an ancilla qubit to $\hat{\xi}$, such that the only qubit dynamics is from the interaction picture Hamiltonian    
$\hat H_{\mathrm{int}} (t) = \frac{1}{2} F(t) \hat{\xi}(t)  \hat \sigma_z$.  We then use the dephasing of the qubit to {\it define} the MGF of the noise in the quantum case, exactly like we would if the noise were classical:
\begin{align}
\label{Eq:MGFqu}
  &   \Lambda[F( t );t_f] \equiv 
         \langle \hat{\sigma}_{-}(t_f) \rangle / \langle \hat{\sigma}_{-}(0) \rangle,
         \\
        % \left. {\frac{\rho_{ \uparrow  \downarrow } (t)} 
        % {\rho_{ \uparrow  \downarrow } (0)} } \right|_{\lambda ,F( t )}  \\
\label{Eq:MGFK}
         = &
        \textrm{Tr} \left[
            \mathcal{T} e^{- \frac{i}{2} \int_0^{t_f} dt' F(t') \hat{\xi}(t')}
            \hat{\rho}_{\rm B}
            \tilde{\mathcal{T}} e^{-\frac{i}{2} \int_0^{t_f} dt' F(t') \hat{\xi}(t')}
        \right].    
        % \left \langle e^{ -i \lambda \int_0^t dt'  \hat B ( t' ) F ( t' )} 
        % \right\rangle_\mathcal{K},
\end{align}
Here $\hat{\rho}_B$ is the initial bath density matrix, the trace is over bath degrees of freedom, and $\mathcal{T}$ ($\tilde{\mathcal{T}}$) is the time-ordering (anti-time-ordering) symbol.  Expanding $\Lambda$ in powers of $F(t')$ defines correlation functions at a given order with a particular time-ordering prescription (the so-called Keldysh ordering). We stress that this approach amounts to trying to ascribe the qubit evolution to an effective classical stochastic process; this correspondence then defines cumulants (and implicitly a quasiprobability) for the quantum noise of interest.  

For truly classical noise, the definition in Eq.~\eqref{Eq:MGFK} reduces to the classical MGF in Eq.~\eqref{Eq:MGFcl}.
%Thus, the Keldysh prescription amounts to using qubit dephasing by quantum noise as a means to {\it define} the appropriately-ordered higher moments of the noise \cite{Levitov1996,Nazarov2003}.  
For quantum baths comprising of harmonic oscillators, and with a noise operator $\hat \xi(t)$ that is linear in bath raising and lowering operators, operator ordering plays no role in the definition of cumulants.  This is because commutators of $\hat \xi(t)$ with itself at different times are numbers, not operators (see Appendix~\ref{AppSec:op.order} for an explicit proof).  As a result, the Keldysh-ordered MGF in Eq.~\eqref{Eq:MGFK} is equivalent to the classical MGF in Eq.~\eqref{Eq:MGFcl} with $\xi(t)$ directly replaced by the quantum noise operator $\hat \xi(t)$.  This is the only kind of bath explicitly discussed in Ref.~\cite{Viola2016} (though the neglect of operator ordering issues is not discussed).   We stress that ignoring operator ordering (i.e., not using the full definition in Eq.~\eqref{Eq:MGFK}) fails in almost any other situation.  In particular, it is not valid for non-Gaussian quantum baths with nonlinear coupling or intrinsic nonlinearity, where the Keldysh ordering in Eq.~\eqref{Eq:MGFK} leads to nontrivial corrections in the quantum noise cumulants (see Appendix~\ref{AppSec:op.order} for further discussion.)

The moments and corresponding quasiprobability defined via Eq.~\eqref{Eq:MGFK} are intrinsic to the noisy system; they predict the outcomes of a wide class of schemes designed to measure this noise
\cite{footnote2}.
%\cite{Nazarov2003,Hofer2017}.  
They also have a direct role in Keldysh non-equilibrium field theory:  they characterize the fluctuations of the ``classical" field associated with the operator $\hat{\xi}(t)$.
This provides an alternate, extremely physical way to understand the Keldysh-ordered cumulants, one that transcends simply viewing this prescription as a formal consequence of expanding interaction-picture operators.
At each order, the Keldysh-ordered correlation function describes the intrinsic fluctuations of the system \cite{Kamenev2011book}.  In contrast, the remaining independent correlation functions at the same order describe how the system responds to external fields which couple to $\hat{\xi}(t)$ 
%(either higher-order response coefficients, or generalized noise susceptibilities; 
(see Appendix~\ref{AppSec:flvsrp} for a complete discussion).
We stress again that at each order,  the Keldysh-ordered correlation function is precisely the correlation function ``seen" by the qubit.  

It follows that the Keldysh-ordered cumulants 
$C^{(k)}(\vec{t}_{k}) \equiv \langle \langle \hat{\xi}\left(  t_1\right) \cdots 
\hat \xi \left(  t_k\right) \rangle\rangle_\mathcal{K}$ of the noise can be generated from   
$\chi[F(t);t_f] \equiv \ln \Lambda[ F(t); t_f ] $ via
\begin{equation}
\label{eq:ncdef}
    \chi[F(t);t_f]  =  \sum_{\ell=1}^\infty\!\frac{(-i)^\ell }{  \ell !}\prod_{j=1}^\ell
    \left[ \int_0^{t_f}\!\! dt_j F(t_j) \right]\!C^{(\ell)}(\vec{t}_{\ell}),
\end{equation}
where we define $\vec{t}_n \equiv (t_1,\ldots,t_n)$.
% , and the dependence of generating functions on the integration time $t_f$ and filter function $F( t )$ is implicitly assumed hereafter. 
Explicit expressions for the first few Keldysh-ordered cumulants are provided in Eqs.~(\ref{AppEq:Kcml_2}) and~(\ref{AppEq:Kcml_3}) of Appendix~\ref{AppSec:exp23cml}.  The Keldysh-ordered second cumulant is simply a symmetrized correlation function, whereas the third cumulant corresponds to suppressing time-orderings where the earliest operator appears in the middle of an expectation value.  

For stationary noise, the $k$th order cumulant $C^{(k)}(\vec{t}_{k})$
only depends on the $k-1$ time separations $\tau_j \equiv t_{j+1}-t_1$, $j= 1,\ldots, k-1 $. 
We define the quantum polyspectra as Fourier transforms of the Keldysh-ordered cumulants:
% with respect to $\{ \tau_j \}$:
\begin{align}
    &S_{n} [\vec{\omega}_{n}] \equiv \!\int_{\mathbb{R}^n}\! d\vec{\tau}_{n} \,
    e^{-i \vec{\omega}_{n}\cdot\vec{\tau}_{n}} C^{(n+1)}(\vec{\tau}_{n}), \quad n \geq 1. 
\label{eq::polyspectra}
\end{align}
% which can in general take complex values. 
For discussion of classical polyspectra, see Refs.~\cite{Elgar1994,Viola2016,Ramon2019}. 
The 
% zero frequency 
$\omega_j \to 0 $ limit
%\quad (j= 1,\ldots, n)$ 
of $S_{n} [\vec{\omega}_{n}]$ characterize fluctuations in $\hat m=\int_0^t dt'  \hat \xi ( t' ) $
in the long-time limit (so-called full counting statistics (FCS)).  
This is the typical setting where the Keldysh approach has found great utility, largely for studying electronic current fluctuations.  Here we extend the method to study non-classical, non-Gaussian noise {\it at non-zero frequencies} (see also Ref.~\cite{PekolaPRB2006} for an application to frequency-dependent current noise).

%%%%%%%%%%%%%%%%%%%%%%%%%%%%%
\begin{figure*}[t]
  \includegraphics[width=180mm]{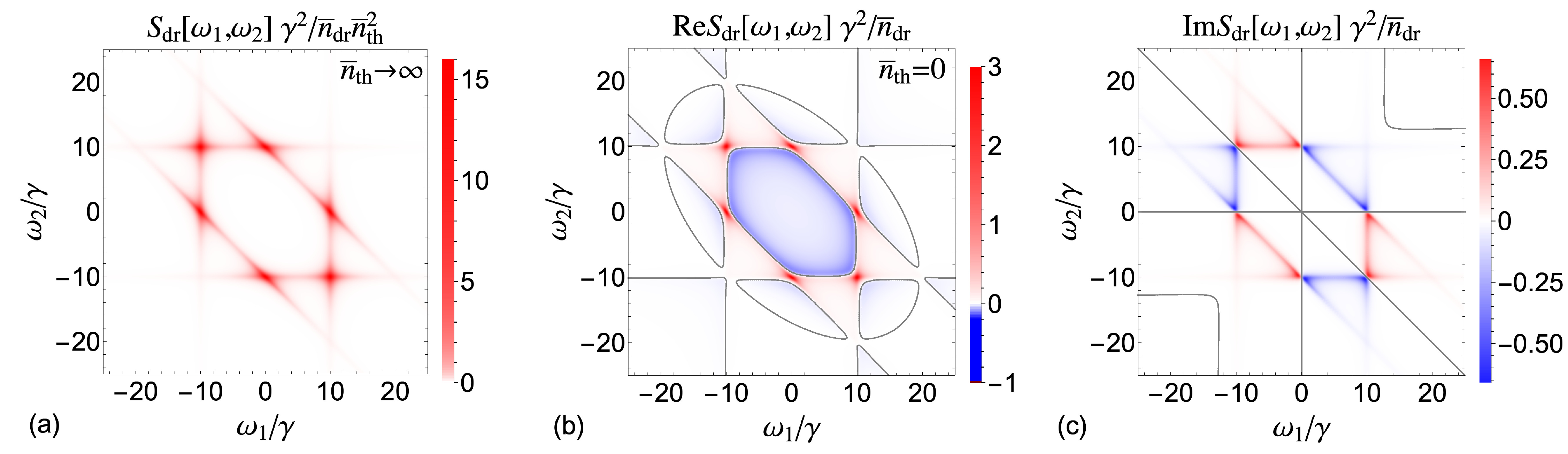}
      \caption{Normalized quantum bispectra of drive-dependent photon fluctuations for large detuning $\delta =10 \gamma$ (a) in the classical limit ${\bar n}_{ \rm{th} }\to \infty$, and (b) real part $\mathrm{Re} S_{\mathrm{dr}} [ {\omega}_{1},{\omega}_{2}  ]$, and (c) imaginary part $\mathrm{Im} S_{\mathrm{dr}} [ {\omega}_{1},{\omega}_{2}  ]$ in the extreme quantum limit ${\bar n}_{ \rm{th} }=0$.
      $\gamma$ is the cavity damping rate, $\bar{n}_{\rm th}$ is the bath thermal number.  
      The red (blue) regions correspond to positive (negative) values, while the gray contour lines indicate zeros.  
    %   The bispectra are normalized by setting $\gamma =1$ and ${\bar n}_{ \rm{dr} }=1$. 
     While the classical-limit bispectrum in (a) is real and positive (as expected from a classical calculation), the quantum-limit bispectrum exhibits negativity in (b) and an imaginary part in (c).
    % The imaginary part $\mathrm{Im} S_{\mathrm{dr}} [ {\omega}_{1},{\omega}_{2}  ]$ is independent of ${\bar n}_{ \rm{th} }$, and thus is fully negligible in (a).
    }
    \label{fig:S2_10}
\end{figure*}
%%%%%%%%%%%%%%%%%%%%%%%%%%%%%

%%%%%%%%%%%%%%%%%%%%%%%%%%%%%%%%%%%%%%%%%%%%%%%%%%%%%%%%%%%%%%%%%%%%%%%%%%%%
%%%%%%%%%%%%%%%%%%%%%%%%%%%%%%%%%%%%%%%%%%%%%%%%%%%%%%%%%%%%%%%%%%%%%%%%%%%%

\section{Quantum noise model}

The utility of our approach can be illustrated by studying a concrete, nontrivial example of quantum non-Gaussian noise: the energy fluctuations of a driven-damped bosonic mode. In what follows, we focus on the physics of the frequency-dependent third cumulant, the so-called quantum bispectrum (QBS) $S_2[\omega_1,\omega_2]$; we drop the subscript $2$ hereafter.
%The quantum non-Gaussian noise of interest will be the energy fluctuations of a driven-damped bosonic mode. 
The QBS reveals a host of physics here that is not manifest in the low-frequency fluctuations (as studied in
\cite{Harris2010,Clerk2011,Clerk2016}).

Our ``bath" here is a driven damped cavity mode $c$ (frequency $\omega_c$, Markovian energy decay rate $\gamma$).  As discussed, the Keldysh-ordered cumulants of the photon-number shot noise can be derived by coupling the number operator $\hat n = \hat c^\dag \hat c$ of the driven cavity to an ancilla qubit via $\hat H_{\mathrm{int}} \left( t\right)=    F(t) \hat n \hat \sigma_z /2$; the desired quantities are then encoded in the qubit coherence via Eq.~\eqref{Eq:MGFqu}. Working in a rotating frame at the drive frequency, and letting $\hat{\rho}$ denote the qubit-cavity reduced density matrix, the system dynamics follows the master equation
\begin{equation}
    \dot {\hat \rho}  =  
    - i [ \hat H_0 + \hat H_{\mathrm{int}} (t) ,\hat \rho ] + \gamma ( {\bar n}_\mathrm{th} + 1 ) \mathcal{D} [ \hat c ]\hat \rho  + \gamma {\bar n}_\mathrm{th} \mathcal{D} [\hat c^\dag ]\hat \rho.
\label{Eq:meq}
\end{equation}
Here $ \mathcal{D} [ \hat A  ]\hat \rho  = \hat A\rho {\hat A^\dag } -  ( {{\hat A^\dag }\hat A\hat \rho  + \hat \rho {\hat A^\dag }\hat A}  )/2$ is the Lindblad dissipator and ${\bar n}_\mathrm{th} $ the thermal photon number associated with the cavity dissipation. The cavity Hamiltonian reads $\hat H_0 =- \delta {\hat c^\dag }\hat c - \left( f \hat c + \mathrm{H.c.} \right)  $, where $f$ ($\delta$) denotes the drive amplitude (detuning).
% Noting that we set $\hbar=1$ in our calculations, the classical limit can be achieved equivalently by taking the high temperature limit ${\bar n}_\mathrm{th} \to \infty $ (see Ref.~\cite{Clerk2011} for a detailed discussion). 

The Keldysh-ordered MGF $\Lambda$ can now be computed by solving the master equation in Eq.~\eqref{Eq:meq}; we stress that the qubit is introduced here as a theoretical tool for extracting the cumulants to appropriately characterize the quantum noise of interest.
Even with an arbitrary time-dependent coupling $F(t)$, the qubit dephasing can be solved exactly using an extension of the phase space method in Ref.~\cite{Utami2007} (see also Appendix~\ref{AppSec:phspmet}). An equivalent approach is to calculate correlation functions using standard techniques (e.g., quantum regression theorem, Heisenberg-Langevin equations)~\cite{Gardiner2004}, and then apply the Keldysh ordering defined in Eqs.~\eqref{Eq:MGFK} and \eqref{eq:ncdef}. In what follows, we will always take the long-time limit $t_f \rightarrow \infty$, making the fluctuations stationary.  
One finds the QBS can be written as:
\begin{align}
    S[\omega_1,\omega_2] =
            S_{\rm th}[\omega_1,\omega_2] + 
            S_{\rm dr}[\omega_1,\omega_2],
            \label{eq:QBSDecomposition}
\end{align}
where the first term is completely independent of the drive $f$, and the second term is proportional to $|f|^2$.

\section{Quantum bispectrum}
\subsection{Drive-independent fluctuations}

The $f$-independent QBS $S_{\rm th}[\omega_1,\omega_2]$ can be calculated by solving Eq.~\eqref{Eq:meq} with $f=0$.  In this case, the cavity relaxes to a thermal steady state with no coherence between different Fock states.  Its fluctuations can thus be mapped to a classical Markovian master equation.  For such a classical and thermal Markov process, the bispectrum $S_{\mathrm{th} }   [ {\omega}_{1},{\omega}_{2}  ] $ 
{\it must} always be real \cite{Semerjian2004,Sinitsyn2016}. 
In Appendix~\ref{AppSec:qbscldr}, we also show that in our case,  $S_{\mathrm{th} }   [ {\omega}_{1},{\omega}_{2}  ] $ must also be positive semidefinite.
Letting $\omega_{3} \equiv -\omega_1 -\omega_2$ in all equations that follow, our full calculation for the Keldysh-ordered QBS yields as expected a real, positive function:
% \begin{equation}
%      S_{\mathrm{th} }   [ {\omega}_{1},{\omega}_{2}  ]  =\mathcal{C}_{{\bar n}_\mathrm{th} }  \gamma^2   \frac{6 \gamma^2+   \sum\limits_{j=1}^3 { \omega}^2 _j  }{ \prod\limits_{j=1}^3 (  \gamma^2  + { \omega}^2 _j   )  }   ,
% \end{equation}
\begin{equation}
\label{Eq:bispmth}
     S_{\mathrm{th} }   [ {\omega}_{1},{\omega}_{2}  ]  =
        \mathcal{C}_{{\bar n}_\mathrm{th} }  \gamma^2   
        \left( 6 \gamma^2+   \sum\limits_{j=1}^3 { \omega}^2 _j  \right) \Big /  \prod\limits_{j=1}^3 (  \gamma^2  + { \omega}^2 _j   ),
\end{equation}
with $\mathcal{C}_{{\bar n}_\mathrm{th} }= {\bar n}_\mathrm{th} ( {\bar n}_\mathrm{th} +1  )  ( 2{\bar n}_\mathrm{th} +1) $.
The frequency dependence of this contribution to the bispectrum is the same both in the classical high-temperature limit ${\bar n}_\mathrm{th} \to \infty $, and in the extreme quantum limit ${\bar n}_\mathrm{th} \to 0$; the only temperature dependence is in the prefactor.  $S_{\mathrm{th} }   [ {\omega}_{1},{\omega}_{2}  ]$ vanishes in the absence of thermal fluctuations (i.e., ${\bar n}_\mathrm{th} \to 0$). 
In the limit ${\bar n}_\mathrm{th} \to 0$, this expression corresponds (as expected) to the bispectrum of asymmetric telegraph noise (see, e.g., Ref.~\cite{Sinitsyn2013}), corresponding to fluctuations between the $n=0$ and $n=1$ Fock states.  
While our general result here suggests that the $\omega$ dependence of the QBS is not sensitive to quantum corrections, we will see that this is not true as soon as a coherent drive is added. 
% to the system.

%%%%%%%%%%%%%%%%%%%%%%%%%%%%%%%%%%%%%%%%%%%%%%%%%%%%%%
%%%%%%%%%%%%%%%%%%%%%%%%%%%%%%%%%%%%%%%%%%%%%%%%%%%%%%

\subsection{Driven fluctuations}
We now consider the drive-dependent
contribution to the bispectrum, $S_{\mathrm{dr} }   [ {\omega}_{1},{\omega}_{2}  ]$
in Eq.~(\ref{eq:QBSDecomposition}).  This quantity only depends on the drive amplitude $f$ through the overall prefactor ${\bar n}_{ \rm{dr} } =4 |f|^2/(\gamma^2+ 4\delta^2)$ (the intracavity photon number generated by $f$).  Note that $S_{\mathrm{dr} }   [ {\omega}_{1},{\omega}_{2}  ]$ remains non-zero at zero temperature, and is the only contribution to the QBS in this limit.    

We find that the drive-dependent QBS shows striking quantum signatures.  In the classical limit of high temperatures, it is always real and positive (similarly to the purely thermal contribution, see Appendix~\ref{AppSec:qbscldr} for detail).  However, as temperature is lowered and quantum fluctuations dominate, this quantity can have a negative real part, and even a non-zero imaginary part.  These 
% purely 
quantum features become more pronounced as the magnitude of the drive detuning $\delta$ is increased.  The real and imaginary parts of $S_{\mathrm{dr} }   [ {\omega}_{1},{\omega}_{2}  ]$ 
are plotted for zero temperature in Figs.~\ref{fig:S2_10}(b) and~\ref{fig:S2_10}(c) for a large drive detuning ($\delta / \gamma =10$).

Consider first the surprising negativity of the real part of the zero-temperature QBS.
Negativity in the zero-frequency limit was already discussed in \cite{Clerk2011,Clerk2016}.  
These works showed that this is a purely quantum effect, and that for large detunings it
makes it impossible
% the magnitude of the third cumulant is so great that the fluctuations cannot be 
to describe the fluctuations by a positive-definite quasiprobability.  Our results show how this striking non-classicality also manifests itself in the {\it non-zero} frequency fluctuations.  
% It also demonstrates that this quantum correction has a non-trivial frequency dependence.  
We find that the QBS $ S_{\mathrm{dr}} [ {\omega}_{1},{\omega}_{2}]$ has a different frequency dependence in the quantum limit (${\bar n}_{ \rm{th} }=0$) versus the classical limit ${\tilde S}_{ \mathrm{cl}}[ {\omega}_{1},{\omega}_{2}]$.  
To see this, we write
% \begin{equation}
%      S_{\mathrm{dr}} [ {\omega}_{1},{\omega}_{2}]  = {\bar n}_{ \rm{dr} }  ( 2{\bar n}_{ \rm{th} }+1)^2 \! \left[ {\tilde S}_{ \mathrm{cl}}[ {\omega}_{1},{\omega}_{2}] + \frac{ {\tilde S}_{ \mathrm{qu}}[ {\omega}_{1},{\omega}_{2}] }{( 2{\bar n}_{ \rm{th} }+1)^2} \right].
% \end{equation}
\begin{equation}
     \frac{S_{\mathrm{dr}} [ {\omega}_{1},{\omega}_{2}] }{{\bar n}_{ \rm{dr} }}
     =   ( 2{\bar n}_{ \rm{th} }+1)^2 \!  {\tilde S}_{ \mathrm{cl}}[ {\omega}_{1},{\omega}_{2}] 
     + 
      {\tilde S}_{ \mathrm{q}}[ {\omega}_{1},{\omega}_{2}]  .
\end{equation}
The first term is the classical contribution which dominates in the high-temperature limit;  ${\tilde S}_{ \mathrm{cl}}[ {\omega}_{1},{\omega}_{2}]$ is  independent of both $ {\bar n}_{ \rm{dr} } ,{\bar n}_{ \rm{th} }$, and is real and positive for all frequencies.  Its form can be found directly from a classical Langevin equation calculation (see Eq.~\eqref{AppEq:QBSdrcl} of Appendix~\ref{AppSec:qbscldr}). 
In contrast, the second term is the temperature-independent quantum correction.  It has a {\it completely different} frequency dependence from the classical limit, as described by ${\tilde S}_{ \mathrm{q}}[ {\omega}_{1},{\omega}_{2}]$
\begin{equation}
    {\tilde S}_{ \mathrm{q}}[ {\omega}_{1},{\omega}_{2}] = -\frac{1}{2 } \! \! \sum\limits_{ \substack{\alpha  \ne \beta \\ \alpha ,\beta  = 1,2,3  }}   \frac{\frac{\gamma }{2} + i{ \omega} _\beta }{ (\gamma  - i { \omega} _\alpha) [( \frac{\gamma }{2} + i { \omega}_\beta  )^2+ \delta ^2]}.
 \label{eq:Sdr2qu}
\end{equation}
This function can have both a negative real part, and a non-zero imaginary part. 
% Any negative region in $\mathrm{Re} S_{\mathrm{dr}}   [ {\omega}_{1},{\omega}_{2}  ]$ is thus solely due to the quantum correction. 
In the quantum limit ${\bar n}_{ \rm{th} }=0$, one finds that real part of the QBS only becomes negative above a critical value of the detuning $|\delta|$.  Moreover, the initial onset of negativity occurs at $\omega_1 = \omega_2 = 0$. In the large-detuning regime $|\delta| \gg \gamma$, the negative region of the QBS is peaked near a polygon
whose shape is defined by the resonance conditions $\omega_j = \pm \delta$ ($j=1,2,3$).

%%%%%%%%%%%%%%%%%%%%%%%%%%%%%
\begin{figure}[t]
  \includegraphics[width=85mm]{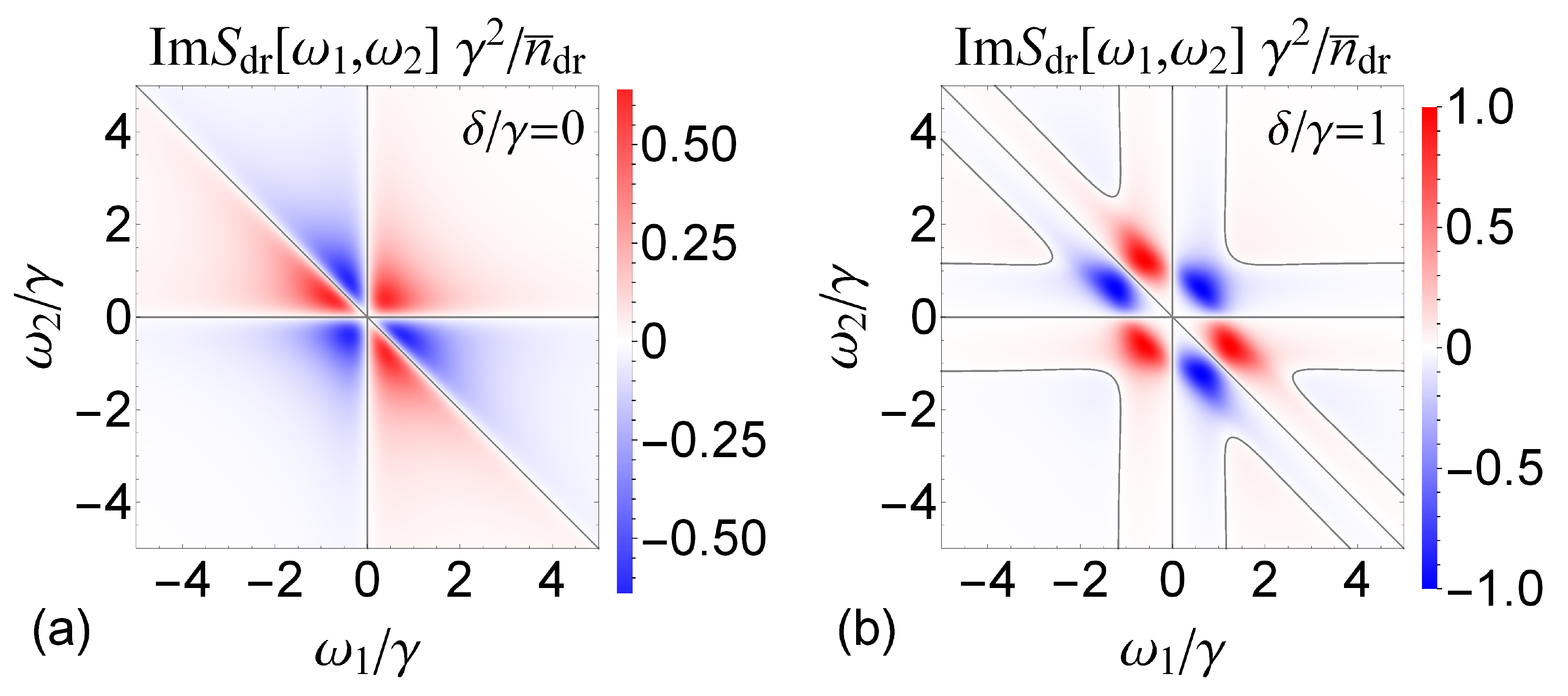}
  \caption{Frequency dependence of the imaginary parts of the
  photon-shot noise bispectrum  $\mathrm{Im} S [ {\omega}_{1} ,{\omega}_{2} ]$  in the extreme quantum limit ${\bar n}_{ \rm{th} }=0$ at different detunings. Parameters: (a) $\delta / \gamma = 0$, (b) $\delta / \gamma = 1$.  Note that for ${\bar n}_{ \rm{th} }=0$ the full quantum bispectrum coincides with the drive-dependent contribution 
  $S_{\rm dr}   [ {\omega}_{1},{\omega}_{2}  ]$.}
    \label{fig:ImS2}
\end{figure}
%%%%%%%%%%%%%%%%%%%%%%%%%%%%%

%%%%%%%%%%%%%%%%%%%%%%%%%%%%%
\begin{figure}[t]
  \includegraphics[width=85mm]{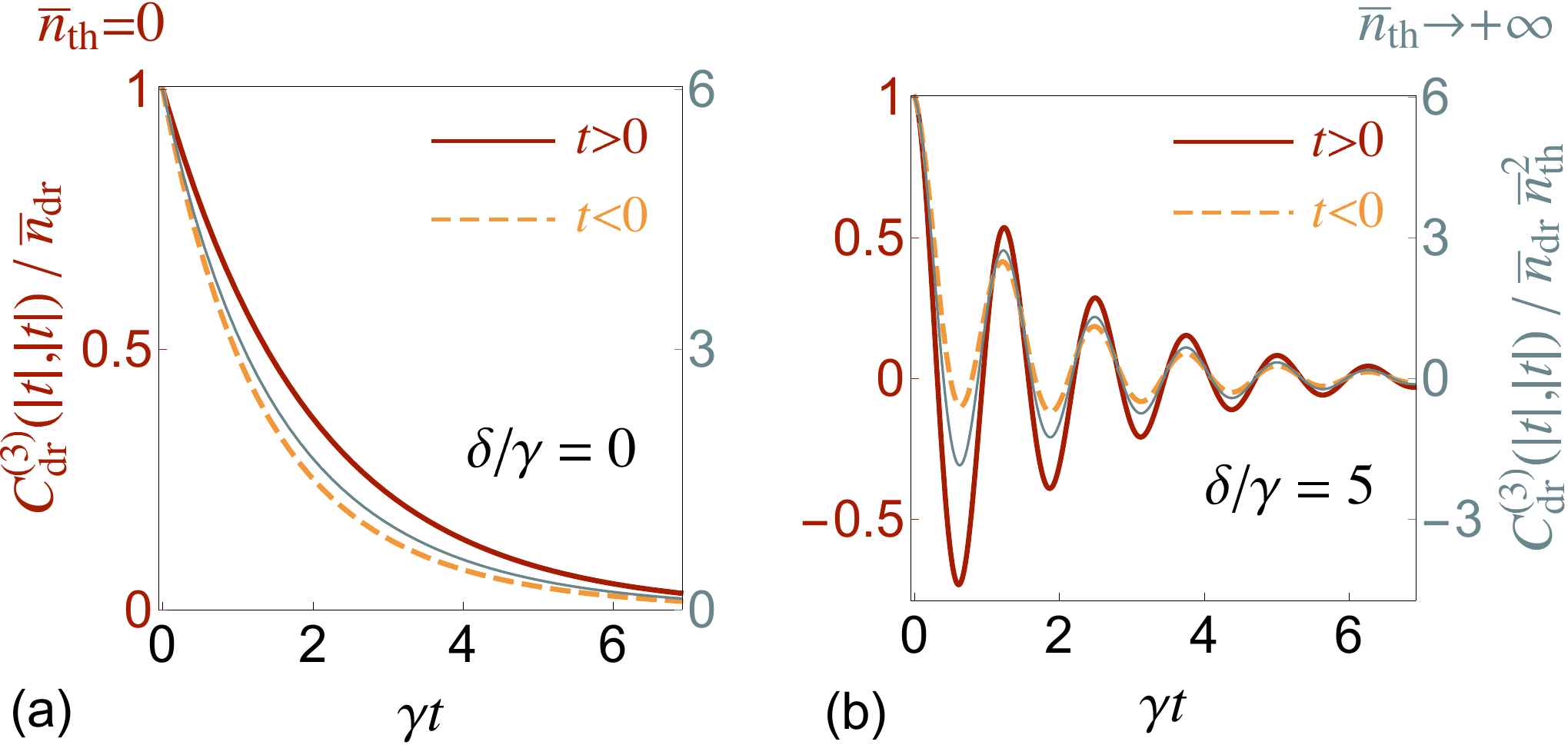}
  \caption{Time-dependent Keldysh-ordered photon-shot noise third cumulant $C_{\mathrm{dr} }^{(3)}(| t|,|t|)$ in the quantum limit ${\bar n}_{ \rm{th} } =0$ for $t>0$ (red solid lines) and $t<0$ (orange dashed lines).  Difference between curves highlights asymmetry  under time reversal $t \to -t$. In contrast, the thin blue curves correspond to the same correlator $C_{\mathrm{dr} }^{(3)}(| t|,|t|)$ in the classical limit normalized by thermal photon number, which is symmetric. All correlation functions are normalized by in-cavity drive photon number ${\bar n}_{ \rm{dr} }$. Detunings are: (a) $\delta=0$, and (b) $\delta/\gamma=5$.}
    \label{fig:timeskn}
\end{figure}
%%%%%%%%%%%%%%%%%%%%%%%%%%%%%

%%%%%%%%%%%%%%%%%%%%%%%%%%%%%%%%%%%%%%%%%%%%%%%%%%%%%%%%%%%%%%%
\section{Imaginary bispectrum and violations of detailed balance}

We now turn to another striking feature of the photon shot noise QBS:  while in the classical, high-temperature limit it is always real, the quantum correction  ${\tilde S}_{ \mathrm{q}}[ {\omega}_{1},{\omega}_{2}]$ has a non-zero imaginary part (see Fig.~\ref{fig:ImS2}). 
This non-trivial imaginary bispectrum can {\it only} be probed at finite frequency:  by its very definition in Eq.~\eqref{eq::polyspectra}, the imaginary part of the QBS
% $\mathrm{Im}  S_{\mathrm{dr}}   [ {\omega}_{1},{\omega}_{2}  ]$   
must vanish if either $ \omega_1=0 $ or $\omega_2 =0$. 
% This is due to the invariance of the time-domain third cumulant $C^{(3)}(\vec{t}_{3}) $ under cyclic permutation of its arguments $\vec{t}_{3}$.

The non-zero imaginary QBS is directly related to the basic symmetries of our quantum noise process, in particular the violation of Onsager-like time symmetry  \cite{Onsager1931,Semerjian2004,Sinitsyn2016}.  
% it reflects the breaking of detailed balance by quantum corrections, and the violation of a generalized Onsager reciprocity relation.  
If a temporal cumulant $C^{(n+1)}(\vec{\tau}_{n})$ is invariant under $\vec{\tau}_{n} \to -\vec{\tau}_{n}$
%  \equiv (-{\tau}_1,\ldots,-{\tau}_{n})$ 
(i.e., the noise process satisfies microscopic reversibility), then the corresponding polyspectrum must be real \cite{Brillinger1967}.
Further, a classical Markov process obeying detailed balance always respects this symmetry.
In our system quantum corrections (as described by ${\tilde S}_{ \mathrm{q}}[ {\omega}_{1},{\omega}_{2}]$) cause a breaking of this symmetry and hence of detailed balance.    
There is a long history of studying detailed balance in 
% non-equilibrium 
driven-dissipative quantum systems (see, e.g., Ref.~\cite{Kubo1966,Agarwal1973,Carmichael1976,Tomita1973,Tomita1974,Carmichael2002}); the QBS provides yet another tool for exploring this physics.  
 In Appendix~\ref{AppSec:sqbathfl}, we discuss another related quantum system which exhibits an apparent breaking of detailed balance, namely a cavity driven by squeezed noise.

For a heuristic understanding of this symmetry breaking, 
we consider a simpler object,  the temporal (Keldysh-ordered) third cumulant $C^{(3)}(\tau_1,\tau_2)$ at $\tau_1 = \tau_2 = t$.  
% This corresponds to a Keldysh-ordered correlation function between $\delta n(t)^2$ and $\delta n(0)$. 
The non-zero imaginary QBS implies that this correlator differs for $t$ and $-t$ (see Fig.~\ref{fig:timeskn}).  Using the definition of Keldysh ordering in Eqs.~(\ref{Eq:MGFK}) and (\ref{eq:ncdef}) we find:
\begin{align}
      \langle \delta \hat n(0) \delta \hat n(t ) \delta \hat n(t )\rangle_\mathcal{K} =  & 
      \frac{1 }{2} \langle \{\delta \hat n(0) ,[\delta \hat n(t )]^2 \} \rangle
      \nonumber \\
     \label{eq:KeldyshOrderedSkewness} 
    &  
  -\frac{ \Theta(-t)}{4}   \langle [ \delta \hat n(t), [ \delta \hat n(t ),\delta \hat n(0)] ] \rangle
  ,
\end{align}
where 
% the subscript $\mathcal{K}$ denotes Keldysh ordering, 
$\delta \hat{n}(t) = \hat{n}(t) - \langle \hat{n}(t) \rangle$, and $\Theta(t)$ is the Heaviside step function.  One finds that any imaginary quantum correction $ \mathrm{Im}  \tilde{S}_q[\omega_1,\omega_2]$ is entirely due to the second term on the RHS; it is thus completely responsible for the lack of time symmetry. 
%One can easily verify that the first symmetrized term is necessarily symmetric in time.  The violation of an Onsager symmetry here is thus entirely due to the Keldysh ordering deviating from the simple symmetrized ordering.

What does this mean physically?  As we have emphasized, the Keldysh ordering is relevant for any measurement protocol that directly probes $\hat{n}(t)$ \cite{Nazarov2003,Hofer2017}.  In contrast, the first term on the RHS of Eq.~(\ref{eq:KeldyshOrderedSkewness}) would be relevant if we correlated a measurement of $\delta \hat{n}$ with a separate, direct measurement of $\delta \hat{n}^2$ (i.e., the Keldysh approach would give this answer for this sort of setup \cite{Hofer2017}).  These protocols are not equivalent:  measuring $\delta \hat{n}$ and then squaring the result has a different backaction than if one directly measured $\delta \hat{n}^2$.  The latter measurement provides less information (and hence has less backaction), as it provides no information on the sign of $\delta \hat{n}$.  This now provides a heuristic way of understanding the second term on the RHS of Eq.~(\ref{eq:KeldyshOrderedSkewness}) (and the consequent lack of time symmetry).  For $t<0$, one is first measuring $\delta \hat{n}^2$.  As a result, the two measurement protocols have different backaction effects, and the two correlation functions are distinct.  In contrast, for $t>0$, the earlier measurement is the same in both protocols, hence the backaction effect is identical, and the two protocols agree.   

While our heuristic explanation here invokes measurement backaction, we stress 
% yet again 
that the Keldysh-ordered correlation function is an {\it intrinsic} property of the driven cavity system \cite{Kamenev2011book,Nazarov2003}, with a relevance that goes beyond the analysis of just a single measurement setup.  Further, this is the ordering that is ``chosen" by our qubit:  if one simply interprets the qubit dephasing as arising from classical noise, then the Keldysh ordered bispectrum  (with its imaginary part) plays the role of the bispectrum of this effective classical noise.

%%%%%%%%%%%%%%%%%%%%%%%%%%%%%%%%%%%%%%%%%%%%%%%%%%%%%%%%%%%%%%%%%%%%%%%%%%%
%%%%%%%%%%%%%%%%%%%%%%%%%%%%%%%%%%%%%%%%%%%%%%%%%%%%%%%%%%%%%%%%%%%%%%%%%%%

%%%%%%%%%%%%%%%%%%%%%%%%%%%%%%%%%%%%%%%%%%%%%%%%%%%%%%%%%%%%%%%%%%%%%%%%%%%%%%%%%%

\section{Conclusions}
We have shown how the Keldysh approach to quantum noise provides a meaningful way to define the polyspectra of non-classical, non-Gaussian noise.  In the experimentally relevant case of photon shot noise fluctuations in a driven-damped resonator, the quantum bispectrum reveals distinct quantum features and a surprising quantum-induced breaking of detailed balance.  We stress that our approach amounts to interpreting the dephasing of a qubit by quantum noise as arising from an effective classical noise process.  As such, the same noise spectroscopy techniques that have been used successfully to measure classical bispectra with qubits \cite{Viola2016,Oliver2019} can be directly used (without modification) to measure our quantum bispectra.  

\section*{Acknowledgements}
We thank L. Viola for useful discussions.  This work was supported as part of the Center for Novel Pathways to Quantum Coherence in Materials, an Energy Frontier Research Center funded by the U.S. Department of Energy, Office of Science, Basic Energy Sciences.

%%%%%%%%%%%%%%%%%%%%%%%%%%%%%%%%%%%%%%%%%%%%%%%%%%%%%%%%%%%%%%%%%%%%%%%%%%%%%%%%%%%%%%%%%%%%%%%%%%
%%%%%%%%%%%%%%%%%%%%%%%%%%%%%%%%%%%%%%%%%%%%%%%%%%%%%%%%%%%%%%%%%%%%%%%%%%%%%%%%%%%%%%%%%%%%%%%%%%
%%%%%%%%%%%%%%%%%%%%%%%%%%%%%%%%%%%%%%%%%%%%%%%%%%%%%%%%%%%%%%%%%%%%%%%%%%%%%%%%%%%%%%%%%%%%%%%%%%

\appendix

%%%%%%%%%%%%%%%%%%%%%%%%%%%%%%%%%%%%%%%%%%%%%%%%%%%%%%%%%%%%%%%%%%%%%%%%%%%%%%%%%%%%%%%%%%%%%%%%%%
\section{When is it necessary to consider Keldysh ordering, and when is it safe to ignore operator ordering in defining quantum noise cumulants?}
\label{AppSec:op.order}

In the main text, we have defined Keldysh-ordered cumulants in Eq.~(\ref{Eq:MGFK}) to characterize a generic quantum noise process; the Keldysh ordering follows directly from standard rules of time evolution in the interaction picture. At first glance, this would seem to contradict the definition in Ref.~\cite{Viola2016}, which directly extends the definition of classical noise cumulants to the quantum case without specifying any particular time ordering of bath operators (see unnumbered equation on page 2 of Ref.~\cite{Viola2016}). Their definition of the $k$-th cumulant can be written as
\begin{align}
C^{(k)}(\vec{t}_{k}) \equiv \langle \langle \hat{\xi}\left(  t_1\right) \cdots \hat \xi \left(  t_k\right) \rangle\rangle,
\label{AppEq:clcml}
\end{align}
where $\langle \langle \cdot \rangle\rangle$ relates the $k$th cumulant to $j$th moments $ \langle \hat{\xi}\left(  t_1\right) \cdots \hat \xi \left(  t_j\right) \rangle $ for $ j \le k$, in exactly the same way as if $\hat{\xi}\left( t \right) $ were classical stochastic variables.  Again, we stress that there is no time-ordering prescription specified here.

As we now show, there is in fact no contradiction between Eq.~(\ref{AppEq:clcml}) and our definition in Eq.~(\ref{Eq:MGFK}).  This is because Ref.~\cite{Viola2016} at the outset restricts their discussion to the specific class of linearly coupled quantum oscillator baths (as is stated explicitly in the introduction of Ref.~\cite{Viola2016}).  We discuss this more in what follows.  Note that while Ref.~\cite{Viola2016} discussed some specific cases where their approach is valid, general conditions for its validity were not provided.  As we show below, the basic requirement is that the commutator of the bath noise operator $\hat{\xi}(t)$ with itself at different times must simply be a number (or more generally, an operator that always commutes with $\hat{\xi}(t)$).  
This is only satisfied if the bath is a collection of harmonic oscillators, and the bath noise operator is linear in mode raising and lowering operators.  We show this explicitly in what follows.

We start with the quantum bath models considered in Ref.~\cite{Viola2016}, which consist of non-interacting bosonic modes $a_k$ with Hamiltonian $H_B$ and noise operator $B(t)$ of the form
\begin{align}
     H_B&=\hbar\sum_k \Omega_k a_k^\dag a_k, 
     \label{AppEq:qoscibathH}\\
     B(t)&=\sum_k (g_k e^{i\Omega_k t} a_k^\dag +\text{h.c.}),
     \label{AppEq:qoscibathop}
\end{align}
where noise operator is linear in raising and lowering operators, and the bath initial state $\rho_B(0)$ is chosen to be diagonal in the Fock basis to ensure stationarity. We now prove that the noise cumulants defined by Eq.~(\ref{AppEq:clcml}), which ignores any operator ordering, agrees with the Keldysh-ordered quantum noise cumulants for these quantum baths. This is equivalent to showing that the Keldysh-ordered moment generating function (MGF) $\Lambda[F( t );t_f]$ in Eq.~(\ref{Eq:MGFK}) now agrees with the MGF without any time ordering
\begin{equation}
     \Lambda_\mathrm{cl} [F( t );t_f] \equiv 
          \textrm{Tr}\left[
              e^{-i \int_0^{t_f} dt' F(t') \hat{\xi}(t')}
            \hat{\rho}_{\rm B}
        \right],
        \label{AppEq:CGFopcl}
\end{equation}
where $\hat{\rho}_B$ is again the initial bath density matrix, and we use $\hat{\xi}(t)$ to denote general bath operators. It is straightforward to see that the cumulants in Eq.~(\ref{AppEq:clcml}) (and in Ref.~\cite{Viola2016}) can be generated by $\ln  \Lambda_\mathrm{cl} [F( t );t_f] $.

We thus seek to prove that $\Lambda_\mathrm{cl} [F( t );t_f] =\Lambda [F( t );t_f] $ for quantum bath described by Eqs.~(\ref{AppEq:qoscibathH}) and (\ref{AppEq:qoscibathop}). Noting that for these baths, the commutators of bath noise operators $[\hat{\xi}(t' ),\hat{\xi}(t'' )]$ are just numbers.  The following identity relations will then hold for generic $F(t)$
\begin{align}
\label{AppEq:Tevolinear}
  &   \mathcal{T} e^{- \frac{i}{2} \int_0^{t_f} dt' F(t') \hat{\xi}(t')} = 
  \exp\left[
  -\hat M_1 (t_f) 
  -\hat M_2 (t_f) 
  %{- \frac{i}{2} \int_0^{t_f}\!\!  dt' F(t') \hat{\xi}(t')}
  %- \frac{1}{8} \int_0^{t_f}\!\!  dt'\!\! \int_0^{t'}\!\!  dt'' F(t') F(t') [\hat{\xi}(t'),\hat{\xi}(t'')]
  \right], \\
\label{AppEq:ATevolinear}
  &   \tilde{\mathcal{T}} e^{-\frac{i}{2} \int_0^{t_f} dt' F(t') \hat{\xi}(t')} =
   \exp\left[
   -\hat M_1 (t_f) 
  +\hat M_2 (t_f) 
  %{- \frac{i}{2} \int_0^{t_f}\!\!  dt' F(t') \hat{\xi}(t')}
  %+ \frac{1}{8} \int_0^{t_f}\!\!  dt'\!\! \int_0^{t'}\!\!  dt'' F(t') F(t') [\hat{\xi}(t'),\hat{\xi}(t'')]
  \right] ,    
  \end{align}
where
\begin{align}
  & \hat M_1 (t_f) =
  \frac{i}{2} \int_0^{t_f}\!\!  dt' F(t') \hat{\xi}(t'), \\
  & \hat M_2 (t_f) =
  \frac{1}{8} \int_0^{t_f}\!\!  dt'\!\! \int_0^{t'}\!\!  dt'' F(t') F(t'') [\hat{\xi}(t'),\hat{\xi}(t'')],
\end{align}
and $[\hat{\xi}(t'),\hat{\xi}(t'')] $ is just a complex-valued function of $t'$ and $t''$. Both equations can be rigorously proved by discretizing the time integral into $N$ infinitesimal time intervals $\delta t = t_f /N$, so that the time- and anti-time-ordered operators can be rewritten as an ordered product of propagators over these time increments, applying the Baker--Campbell--Hausdorff formula, and then taking the continuum limit $\delta t \to 0$~\cite{Plebanski1969,Gardiner2004}. Substituting Eqs.~(\ref{AppEq:Tevolinear}) and (\ref{AppEq:ATevolinear}) into Eq.~(\ref{Eq:MGFK}) for the Keldysh-ordered MGF $\Lambda [F( t );t_f] $, we obtain
\begin{align}
&    \Lambda [F( t );t_f] 
        \nonumber \\
        =&
        \textrm{Tr} \left[
            \mathcal{T} e^{- \frac{i}{2} \int_0^{t_f} dt' F(t') \hat{\xi}(t')}
            \hat{\rho}_{\rm B}
            \tilde{\mathcal{T}} e^{-\frac{i}{2} \int_0^{t_f} dt' F(t') \hat{\xi}(t')}
        \right]  
        \nonumber \\
 =& \textrm{Tr}\left[
              e^{- 2 \hat M_1 (t_f)}
            \hat{\rho}_{\rm B}
        \right] 
    =\textrm{Tr}\left[
              e^{-i \int_0^{t_f} dt' F(t') \hat{\xi}(t')}
            \hat{\rho}_{\rm B}
        \right] 
        \nonumber \\
 =& \Lambda_\mathrm{cl} [F( t );t_f]  ,
 \label{AppEq:qCGFeqcl}
\end{align}
which completes our proof. Incidentally, Eqs.~(\ref{AppEq:Tevolinear}) and (\ref{AppEq:ATevolinear}) will also hold if the commutators $[\hat{\xi}(t'),\hat{\xi}(t'')] $ are still operators, but always commute with the bath operator $\hat{\xi}(t )$ at all times; for this scenario, the commutators $[\hat{\xi}(t'),\hat{\xi}(t'')] $ can be viewed equivalently as numbers as far as dynamics is concerned.

When the bath noise operator is given by Eq.~(\ref{AppEq:qoscibathop}), the bath dynamics will be completely linear, and any nontrivial non-Gaussianity can only be introduced via a non-Gaussian initial state. For example, Ref.~\cite{Viola2016} considered an initial bath state $\rho_B(0)=\rho_{T_1}/2+\rho_{T_2}/2$ as a classical mixture of two thermal states at different temperatures $T_{1}$ and $T_{2}$. The non-Gaussian statistics here can be viewed as a result of the classical uncertainty in two different Gaussian distributions (i.e.~uncertainty in temperature).

We also note that for any quantum bath where Eq.~(\ref{AppEq:qCGFeqcl}) is not true, Keldysh ordering cannot be ignored when defining noise cumulants.   Further, there exist a variety of physical quantum baths where operator ordering plays an important role, and the Keldysh ordering leads to nontrivial corrections in non-Gaussian noise cumulants:
\begin{itemize}
\item{Harmonic oscillator bath, where the bath operator is not linear in raising and lowering operators of the bosonic modes, e.g., the photon shot noise considered in the main text. Here the nonlinearity in the system-bath interaction induces nontrivial non-Gaussian statistics with distinct quantum features.}
\item{Interacting oscillator bath, i.e., bosonic bath with nonlinear dynamics, where the bath operator is linear in raising and lowering operators of the bosonic modes; this includes phonon bath with interactions. Keldysh ordering matters here due to the inherently nonlinear dynamics of the bath.}
\item{Spin bath that exhibits non-Gaussian fluctuations. In this case, both the bath dynamics and the bath operator can induce non-Gaussian statistics, and it is in general nontrivial to apply the Keldysh ordering.}
\end{itemize}
At the formal level, the Keldysh ordering is essential in these cases because the commutator $[\hat{\xi}(t' ),\hat{\xi}(t'' )]$ between bath operators at different times is a nontrivial operator (i.e., nonzero and does not commute with  $\hat{\xi}(t )$). Due to the existence of these realistic examples of noise models where operator ordering is nontrivial, depending on the nature of quantum environments of interest, it may be important to be aware of the distinction between the most generic definition in Eq.~(\ref{Eq:MGFK}), incorporating the Keldysh ordering, and the special case of linearly coupled oscillator baths, where Eq.~(\ref{AppEq:CGFopcl}) applies and the ambiguity in operator ordering can be ignored. In the main text, we also provide a concrete example where the Keldysh ordering results in unique quantum features in the quantum bispectrum, revealing a surprising breaking of detailed balance due to quantum fluctuations.

%%%%%%%%%%%%%%%%%%%%%%%%%%%%%%%%%%%%%%%%%%%%%%%%%%%%%%%%%
%%%%%%%%%%%%%%%%%%%%%%%%%%%%%%%%%%%%%%%%%%%%%%%%%%%%%%%%%
\section{Keldysh-ordered quasiprobability distribution as a description of intrinsic noise}
\label{AppSec:mass}

In the main text, we have focused on using the quantum bispectrum to understand the physics of the nontrivial energy fluctuations in a driven damped harmonic oscillator, and we state that a quasiprobability distribution can be defined for the Keldysh-ordered moment generating function (MGF) $\Lambda$. To elaborate on this and illustrate the generality of the Keldysh approach, here we briefly summarize another paradigmatic measurement setup, where the Keldysh-ordered quasiprobability distribution explicitly determines the measurement result. For more detailed discussions, the reader can refer to Refs.~\cite{Nazarov2003,Clerk2011,Hofer2017}.

Nazarov and Kindermann~\cite{Nazarov2003} considered an idealized setup for measuring the statistics of a generic quantum observable $\hat \phi [F( t );t_f]  = \int_0^{t_f} dt' F(t') \hat{\xi}(t')$
making use of an infinitely heavy mass. Without loss of generality, we assume that the detector mass is moving in $1$-dimensional space with the Hamiltonian $\hat H_{\mathrm{bath}} = \hat V (\hat x) + \hat p^2 / 2m$. We also take the limit where the detector mass is infinitely heavy, i.e., $m \to \infty$, to avoid classical back action of the detector, so that it only measures \textit{fluctuation} properties of the bath. The bath operator $\hat{\xi}(t) $ is coupled to the detector mass via the position operator $\hat x$, described by the Hamiltonian
\begin{equation}
     \hat H_{\mathrm{int}} \left( t\right) 
     =  F(t) \hat{\xi}(t) \hat x, 
\end{equation}
and the detector-bath coupling $\hat H_{\mathrm{int}}  ( t )$ is on for time $t_f$.

If we were measuring a classical variable $ {\xi}(t)$, the net effect of the coupling would be to simply shift the detector momentum by an amount $ \phi  = \int_0^{t_f} dt' F(t')  {\xi}(t')$. For a classical stochastic process $ {\xi}(t)$, the final momentum probability distribution function of the detector is just given by a convolution of the initial momentum distribution, and the probability distribution of momentum shifts $P(\phi)$. In the quantum regime, the detector state can no longer be represented by a classical probability distribution, but the aforementioned physical intuition still applies to a quasiprobability distribution, i.e., the Wigner function $W(x,p)$ of the detector state. However, for an operator $\hat{\xi}(t)$, the classical probability distribution $P(\phi)$ should be replaced by a Keldysh-ordered quasiprobability distribution $P( \phi ; x)$, which is dependent on the detector position $x$. The Wigner function of the detector at final time $t_f$ can thus be written as
\begin{equation}
     W(x,p; t_f) 
     =  \int d\phi 
     P( \phi ; x) 
     W(x,p - \phi; 0) , 
\end{equation}
which reproduces Eq.~(15) in Ref.~\cite{Nazarov2003}. The Keldysh-ordered MGF discussed in Eq.~(\ref{Eq:MGFK}) in the main text directly characterizes the quasiprobabilities $P( \phi ; x)$. The idealized measurement here can be viewed as an illustration of the fact that Keldysh-ordered noise cumulants are intrinsic properties of the quantum bath, characterizing fluctuation properties. In the following section, we will also provide a rigorous justification, making use of the path-integral formulation of the Keldysh technique.
%%%%%%%%%%%%%%%%%%%%%%%%%%%%%%%%%%%%%%%%%%%%%%%%%%%%%%%%%
%%%%%%%%%%%%%%%%%%%%%%%%%%%%%%%%%%%%%%%%%%%%%%%%%%%%%%%%%

%%%%%%%%%%%%%%%%%%%%%%%%%%%%%%%%%%%%%%%%%%%%%%%%%%%%%%%%%%%%%%%%%%%%%%%%%%%%%%%%%%%%%%%%%%%%%%%%%%
\section{Distinguishing fluctuations from response properties}
\label{AppSec:flvsrp}

As discussed in the main text, a general $n$-point quantum correlation function describes both the intrinsic fluctuation properties of the system of interest (i.e., quantities that play the role of classical noise), as well as the response properties of the system to external applied fields.  The situation is very clear at second order, where the product $\hxi(t) \hxi(t')$ can be decomposed as the sum of a commutator and an anti-commutator.  The commutator determines the retarded Green function
\begin{equation}
   G^R(t) \equiv -i \Theta(t) \left \langle [ \hxi(t), \hxi(0) ] \right \rangle.
\end{equation}
This describes how the average value $\langle \hxi(t) \rangle$ changes to first order in response to an external perturbing field $V(t)$ entering the Hamiltonian as 
\begin{equation}
    \hat{H}_{\rm ext}(t) = V(t) \hxi. 
\end{equation}
The relevant Kubo formula is:
\begin{equation}
    \delta \langle \hxi(t) \rangle = 
        \int_{-\infty}^{\infty} dt' G^R(t-t') V(t').
\end{equation}
In contrast, the anti-commutator describes the symmetrized noise spectral density:
\begin{equation}
    S[\omega] \equiv \frac{1}{2} \int dt e^{i \omega t} \langle \{ \hxi(t), \hxi(0) \} \rangle .
\end{equation}
As has been discussed in many places (see, e.g., Ref.~\cite{ClerkRMP}), this spectral density plays the role of a classical noise spectral density.  

The Keldysh technique provides an unambiguous way of extending this separation between noise and response to higher orders.  A full exposition of this method is beyond the scope of this paper; we refer the reader to Ref.~\cite{Kamenev2011book}.  We sketch the main ideas needed here.  In the path-integral formulation of the Keldysh technique, each operator corresponds to two different fields, the classical field $\xi_{\rm cl}(t)$ and the quantum field $\xi_{\rm q}(t)$.  Averages of these fields (weighted by the appropriate Keldysh action describing the system) then correspond to operator averages with a particular time ordering.  One finds that:
\begin{itemize}
    \item
        Averages only involving quantum fields are necessarily zero.
    \item 
        Averages involving at least one classical field $\xi_{\rm cl}(t)$ and one or more quantum fields $\xi_{\rm q}(t)$ can always be interpreted as response coefficients to an external perturbation 
        of the form $\hat{H}_{\rm ext}(t)$.
    \item
        Averages {\it only} involving classical fields $\xi_{\rm cl}(t)$ do not correspond to any kind of response function.  Instead, they describe the intrinsic fluctuation properties of the system
\end{itemize}
Formally, this dichotomy arises because the perturbation $\hat{H}_{\rm ext}(t)$ enters the action of the system as a term that {\it only} involves the quantum field, i.e., $S_{\rm ext} = \int dt V(t) \xi_{\rm q}(t)$.  Perturbation theory in $V(t)$ thus necessarily introduces powers of the quantum field.  
For example, at second order we have:
\begin{itemize}
    \item 
        The average $\overline{\xi_{\rm cl}(t) \xi_{\rm q}(0)}$ is directly proportional to the retarded Green function $G^R(t)$, and thus describes linear response to the external field.
    \item
         The average $\overline{\xi_{\rm cl}(t) \xi_{\rm cl}(0)}$ is proportional to
         $\langle \{ \hxi(t), \hxi(0) \} \rangle$ and thus determines the usual symmetrized noise spectral density. 
\end{itemize}

The same decomposition applies at higher orders.  Consider third order correlators.  The average of three classical fields $\overline{\xi_{\rm cl}(t_1) \xi_{\rm cl}(t_2) \xi_{\rm cl}(0)}$ is precisely the Keldysh ordered correlator discussed in the main text; it cannot be associated with a response coefficient.  The remaining non-zero correlators describe different kinds of response:
\begin{itemize}
    \item 
        The average $\overline{\xi_{\rm cl}(t) \xi_{\rm q}(t') \xi_{\rm q}(t'')}$ represents a second-order Kubo response coefficient.  It determines to second order how $\langle \hxi(t) \rangle$ is modified by $\hat{H}_{\rm ext}(t')$ at earlier times (i.e., how it depends on $V(t')$ and $V(t'')$).
    \item
         The average $\overline{\xi_{\rm cl}(t) \xi_{\rm cl}(t') \xi_{\rm q}(t'')}$ describes a first order noise-susceptibility~\cite{Reulet2007}.  It determines how the symmetrized correlator  $\langle \{ \hxi(t), \hxi(t') \} \rangle$ is modified to first order by $\hat{H}_{\rm ext}(t'')$.
 \end{itemize}

The arguments sketched here provided perhaps the deepest justification for considering Keldysh ordered correlation functions: they provide a clear and unambiguous way to distinguish fluctuation properties from response properties.  We stress that an arbitrary correlation function could always be written as a linear combination of the Keldysh-ordered correlator (which describes pure noise) and additional terms describing response properties.

%%%%%%%%%%%%%%%%%%%%%%%%%%%%%%%%%%%%%%%%%%%%%%%%%%%%%%%%%%%%%%%%%%%%%%%%%%%%%%%%%%%%%%%%%%%%%%%%%%
\section{Explicit expressions for the second and third Keldysh-ordered cumulants}
\label{AppSec:exp23cml}

For concreteness, here we provide explicit expressions for the first few Keldysh-ordered cumulants $C^{(k)}(\vec{t}_{k}) \equiv \langle \langle \hat{\xi}\left(  t_1\right) \cdots \hat \xi \left(  t_k\right) \rangle\rangle_\mathcal{K}$ defined by Eqs.~\eqref{Eq:MGFK} and \eqref{eq:ncdef} in the main text. The second order cumulant function $C^{(2)}(\vec{t}_{2}) $ is just the auto-correlation function of $ \hat \xi (  t )$
\begin{equation}
\label{AppEq:Kcml_2}
C^{(2)}(\vec{t}_{2})  =    \langle \langle \hat \xi\left( {t_1} \right)  \hat \xi\left( {t_2} \right) \rangle \rangle_\mathcal{K} =  \frac{1}{2}\langle\{ \delta \hat \xi\left( {t_1} \right) , \delta \hat \xi\left( {t_2} \right) \}\rangle,
\end{equation}
where $\delta \hat \xi=\hat \xi -\langle \hat \xi \rangle  $. However, the third cumulant corresponds to a more complex ordering
\begin{align}
\label{AppEq:Kcml_3}
C^{(3)}(\vec{t}_{3}) & =    \langle \langle 
    \hat \xi\left(  {t_1}  \right)  
    \hat \xi\left( {t_2}  \right) 
    \hat \xi\left(  {t_3} \right)
    \rangle \rangle_\mathcal{K} 
\nonumber \\
& =  \frac{1}{4}
    \sum_{\vec{\pi}_3 \in \mathcal{P}_3} 
    K(t_{\pi _1} , t_{\pi _2}, t_{\pi _3})
    %[1-\Theta( t_{\pi _1} - t_{\pi _2}) \Theta( t_{\pi _3} - t_{\pi _2})] 
    \langle \delta \hat \xi (t_{\pi _1}) \delta \hat \xi (t_{\pi _2}) \delta \hat \xi (t_{\pi _3})  \rangle,
    \\
 K(\vec{t}_{3}) & =
    1-\Theta( t_1 - t_2 ) \Theta( t_3 - t_2 ),
\end{align}
where $\mathcal{P}_3$ denotes the set of all possible permutations of $(123)$ indices, and $\Theta(t)$ is the Heaviside step function. Such ordering is given by an average over all permutations of the three displaced operators $\delta \hat \xi (t_j)$, except for the terms where the earliest time appears in the middle position (as implied by the step functions), in agreement with expansion of the operator in Eq.~\eqref{Eq:MGFK} in powers of coupling $F(t')$. A similar expression of Keldysh-ordered third cumulant has also been derived for current operators in Ref.~\cite{PekolaPRB2006}.

%%%%%%%%%%%%%%%%%%%%%%%%%%%%%%%%%%%%%%%%%%%%%%%%%%%%%%%%%%%%%%%%%%%%%%%%%%%%%%%%%%%%%%%%%%%%%%%%%%
\section{Phase space method for computing Keldysh-ordered cumulants of a driven damped cavity}
\label{AppSec:phspmet}

In this section, we outline the phase space method to calculate Keldysh-ordered cumulants. However, we remark that once we have defined the unique Keldysh ordering for each higher cumulant using Eqs.~\eqref{Eq:MGFK} and \eqref{eq:ncdef}, standard techniques for computing multi-point correlation functions (e.g. Langevin equations of motion, and quantum regression theorem) work equally well for the Keldysh-ordered cumulants.

In the phase space method, we need to solve the time evolution of qubit coherence operator $\hat \rho _{ \uparrow  \downarrow } (t) \equiv \langle\uparrow \! |\hat \rho (t)|\! \downarrow \rangle$, so that the qubit coherence can be computed as $\langle \hat{\sigma}_{-}( t ) \rangle = \textrm{Tr} [\hat \rho _{ \uparrow  \downarrow } (t) ]$.
We first restrict to the qubit off-diagonal block 
%$|\! \downarrow \rangle \langle\uparrow \! |$ 
of the master equation in Eq.~(\ref{Eq:meq}) in the main text as
\begin{align}
\dot {\hat \rho}_{ \uparrow  \downarrow }  =  
   &  - i [ \hat H_0  ,\hat \rho_{ \uparrow  \downarrow } ]  
    - i \frac{\lambda}{2} \{  F(t) \hat n   ,\hat \rho_{ \uparrow  \downarrow } \}
    \nonumber \\
 & \,
    + \gamma ( {\bar n}_\mathrm{th} + 1 ) \mathcal{D} [ \hat c ]\hat \rho_{ \uparrow  \downarrow }  + \gamma {\bar n}_\mathrm{th} \mathcal{D} [\hat c^\dag ]\hat \rho_{ \uparrow  \downarrow },
\label{AppEq:meqcoh}
\end{align}
which is a direct extension, with a time modulation $F(t)$ in interaction $\hat H_{\mathrm{int}} \left( t\right)=  {\lambda} F(t) \hat n \hat \sigma_z /2$, of the technique used in Ref.~\cite{Utami2007}. Here we use a constant coefficient $\lambda$ to keep track of orders in expansion on the coupling; by the end of the calculation, one can always set $\lambda=1$. We stress that if we replace the time-independent coupling $\lambda$ with a time-dependent one, the relevant derivations in Ref.~\cite{Utami2007} still hold rigorously, and we refer interested readers to this paper for more detail.

Without loss of generality, the system initial state can be chosen as a product state between the qubit and the cavity, with the cavity in thermal equilibrium. Thus, Wigner function $W(x,p;t)$ of the coherence operator $\hat \rho _{ \uparrow  \downarrow } (t) $ is Gaussian throughout time evolution. Moreover, for the Fourier transform of $W(x,p;t)$, we can assume the following ansatz~\cite{Utami2007}
\begin{align}
&W\left[ {k,q;t} \right] \nonumber \\
 = & e^{ - \nu ( t  )} 
    \exp \left(   - i [ {k\bar x (t) + q\bar p (t)}  ] - \frac{1}{2} (k^2+q^2) \sigma _s (t) \right),
\end{align}
from which the moment generating function can be computed as $\Lambda[F( t );t_f] = e^{-\nu ( t_f )} $. Substituting this ansatz into the master equation in Eq.~\eqref{AppEq:meqcoh}, we then need to solve a set of ordinary differential equations for the coefficient functions
\begin{subequations}
\label{AppEq:cohODE}
\begin{align}
&\dot \nu_\mathrm{th}  =    i\lambda F ( t ) \left( {\sigma _s- \frac{1}{2}} \right) ,\\
&  {\dot \sigma }_s  = \gamma \left( {\bar n}_\mathrm{th}+ \frac{1}{2}  \right) - \gamma {\sigma _s}   - iF ( t ) \lambda \sigma _s^2 + \frac{ i\lambda F \left( t\right)  }{4} , \\
&\dot \nu_\mathrm{dr}  =     \frac{{i\lambda }}{2} F(t)  ({\bar x}^2 +{\bar p}^2)  ,\\
& \dot {\bar x} =  - \delta \bar p + \sqrt 2 \, \mathrm{Im}f  - iF ( t ) \lambda {\sigma _s}  \bar x- \frac{\gamma }{2}\bar x , \label{AppEq:mfGSantcoef-xb}  \\
&  \dot {\bar p}  = \delta \bar x + \sqrt 2 \, \mathrm{Re}f  - i F (t) \lambda {\sigma _s} {\bar p} - \frac{\gamma }{2}\bar p , \label{AppEq:mfGSantcoef-pb} 
  \end{align}
\end{subequations}
where the exponent $\nu (t)=\nu_\mathrm{th}(t) +\nu_\mathrm{dr}(t)$ can be written as a sum of drive-independent and drive-dependent parts.

The Keldysh-ordered cumulants $C^{(\ell)}(\vec{t}_{\ell})$ can now be extracted using the equation (see Eq.~\eqref{eq:ncdef} in the main text)
\begin{equation}
\label{AppEq:ncdef}
    \nu (t_f)  = - \sum_{\ell=1}^\infty\!\lambda^\ell\frac{ (-i)^\ell }{  \ell !}\prod_{j=1}^\ell
    \left[ \int_0^{t_f}\!\! dt_j F(t_j) \right]\!C^{(\ell)}(\vec{t}_{\ell}),
\end{equation}
i.e., the cumulants can be obtained by solving Eqs.~\eqref{AppEq:cohODE} perturbatively in orders of $\lambda$, and comparing the results to the integrals above. Since the cumulant functions $C^{(\ell)}(\vec{t}_{\ell})$ must be symmetric over permutations of its variables $\{\vec{t}_{\ell} \}$, such procedure will lead to a unique result. For example, for the photon shot noise in a driven damped cavity discussed in the main text, first few drive-independent contributions to cumulants are given by
\begin{subequations}
\begin{align}
 C_{\mathrm{th} }^{(1)}( t_1)
    = & {\bar n}_\mathrm{th}  ,\\
 C_{\mathrm{th} }^{(2)}(\vec{t}_2) 
    = & {\bar n}_\mathrm{th} ( {\bar n}_\mathrm{th} +1 ) e^{-\gamma | t_1 - t_2 |}, \\
 C_{\mathrm{th} }^{(3)}( \vec{t}_{3}) 
    = & {\bar n}_\mathrm{th} ( {\bar n}_\mathrm{th} +1 )  ( 2{\bar n}_\mathrm{th} +1 ) \nonumber \\
& \times
    \exp \left(  - \frac{\gamma }{2} | t_1 - t_2 | - \frac{\gamma }{2}| t_2 - t_3 | - \frac{\gamma }{2}| t_1 - t_3 |  \right) . \label{AppEq:cmlth_3} 
  \end{align}
\end{subequations}
Taking Fourier transform of Eq.~\eqref{AppEq:cmlth_3} for the third cumulant, we obtain the drive-independent QBS, as given by Eq.~\eqref{Eq:bispmth} in the main text.

\section{Quantum bispectrum (QBS) probed by qubit dephasing}

In the main text and above, we introduced the ancilla qubit mostly as a theoretical tool to characterize the quantum bath fluctuations. However, as mentioned in the main text, the qubit-bath system is also a well-studied experimental probe to measure the QBS of a given quantum bath. The QBS of the bath can be extracted, by measuring the qubit coherence function $\langle \hat{\sigma}_{-}(t_f) \rangle $ evolving under given filter functions $F(t)$. Ref.~\cite{Viola2016} discusses a systematic approach to reconstruct the bispectrum using this technique of qubit noise spectroscopy.

\begin{figure}[t]
  \includegraphics[width=85mm]{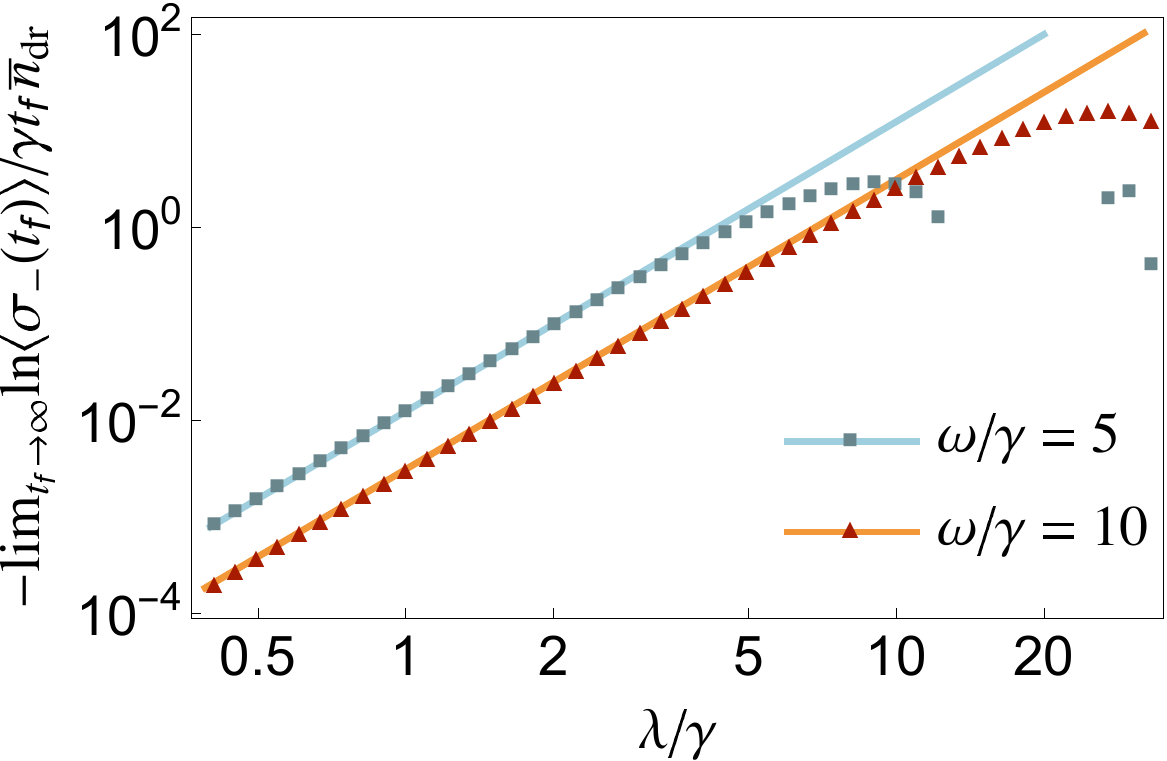}
  \caption{Photon shot noise induced qubit frequency renormalization, as defined from the long time limit of the qubit coherence function, $-\lim_{t_f \to \infty} 
  \text{Im} \ln \langle \hat{\sigma}_{-}(t_f) \rangle/t_f $.  We use here a filter function $F(t) =  \lambda (\sin 2 \omega t + \cos \omega t ) $, and plot the frequency shift as function of coupling strength $\lambda$.  The two curves correspond to two different choices of filter function center frequencies $\omega$ (as indicated in the legend).  Solid lines depict contributions from QBS and have exact slopes of 3, as these terms are proportional to $\lambda^3$, whereas the data points are exact results by solving Eqs.~\eqref{AppEq:cohODE} numerically (and thus include contributions from all higher-order odd noise cumulants). The QBS prediction describes the induced phase shift accurately over a range of weak to moderate couplings. Data points with an opposite frequency shift are not shown (the QBS approximation completely breaks down for these points). Parameters: $\delta =10 \gamma$, ${\bar n}_\mathrm{th} =0 $, ${\bar n}_\mathrm{dr} =1 $.}
    \label{figApp:qubitdep}
\end{figure}

In this section, we apply this idea to the specific noise model discussed in the main text. We consider qubit dephasing due to photon shot noise of a driven damped cavity mode, as described by the master equation in Eq.~(\ref{Eq:meq}). As an illustration, we focus on the idealized filter function
\begin{align}
F(t) = 
    \lambda (\sin 2 \omega t + \cos \omega t ) ,
     \quad
      t \in [0, t_f],
\end{align}
where $\lambda$ characterizes the coupling strength. This filter function is chosen such that for any coupling strength, the qubit coherence has no dependence on the real part of the QBS, i.e.,
\begin{align}
\text{Im} \ln \langle \hat{\sigma}_{-}(t_f) \rangle
   & =  \text{Im} \chi[F(t);t_f] 
    \nonumber \\
   & =\frac{\lambda^3  t_f}{16} 
    \text{Im}  S[\omega,\omega]
    +o(\lambda^5 ) .
\end{align}
As discussed in the main text, the imaginary part of the QBS, which can be computed from Eq.~(\ref{eq:Sdr2qu}), is a unique quantum feature and only depends on driven fluctuations. This phase shift is solely due to the non-Gaussian noise cumulants, and will be absent if we treat the noise operator as Gaussian.
We compare above prediction based on the QBS to the induced frequency shift in the exact qubit coherence function in the long-time limit
\begin{align}
  \lim_{t_f \to \infty} 
  \frac{\text{Im} \ln \langle \hat{\sigma}_{-}(t_f) \rangle}{t_f} ,
\end{align}
which is calculated numerically by solving Eqs.~\eqref{AppEq:cohODE}. The results are plotted in Fig.~\ref{figApp:qubitdep} for the case of zero temperature ${\bar n}_\mathrm{th} =0 $, where the qubit dephasing is solely due to driven fluctuations. As shown in the plot, the QBS prediction agrees excellently with the exact result for small coupling $\lambda$ as expected, but will deviate from the exact result as coupling $\lambda$ increases. The QBS prediction works even at moderate couplings $\lambda / \gamma \sim 1$, because the higher cumulants here are suppressed by the large detuning $\delta / \gamma \gg 1$.  The QBS thus has a concrete operational interpretation: it quantifies the leading order non-Gaussian correction in the qubit dephasing due to a given quantum noise process $\hat{\xi}(t) $.

%%%%%%%%%%%%%%%%%%%%%%%%%%%%%
%%%%%%%%%%%%%%%%%%%%%%%%%%%%%

\section{Proof of non-negative energy shot noise bispectrum in a classical driven damped oscillator}
\label{AppSec:qbscldr}

In the classical limit ${\bar n}_\mathrm{th} \to \infty $, the cavity mode annihilation operator $\hat c$ in the main text can be described by a classical stochastic variable $c(t)$, describing the amplitude of a driven damped classical harmonic oscillator. The equation of motion is now given by
\begin{equation}
d c  =   - ( {\gamma }/{2}-i\delta )c dt + if dt+ \sqrt{ \gamma{\bar n}_\mathrm{eff} }  dW ,
\end{equation}
where ${\bar n}_\mathrm{eff} ={\bar n}_\mathrm{th} +1/2 \simeq {\bar n}_\mathrm{th}$ (the $1/2$ correction is added so that second-order correlators between $c(t)$, $c^*(t')$ match their symmetrized quantum counterparts), and $dW$ is a complex-valued Wiener increment. The solution to this stochastic differential equation can be written as $c(t) = c_0 + \zeta (t)$, where $c_0 = \overline{ c(t) } $ is a complex constant number, and $\zeta (t)$ is a complex zero-mean stochastic variable. In the long-time limit, $\zeta (t)$ is Gaussian and stationary, satisfying the equation
\begin{equation}
\overline{\zeta^*(t)\zeta(t')}
    ={\bar n}_\mathrm{eff}  \exp \!\left[-i \delta (t-t') -\frac{\gamma}{2} |t-t'|\right],
\end{equation}
whereas all other second correlators vanish $\overline{\zeta(t)\zeta(t') }   = [ \overline{\zeta^*(t)\zeta^*(t') } ]^*   \equiv 0 $. The photon number operator $\hat n$ then corresponds to the energy of the classical oscillator $n (t)=|c(t)|^2$, so that its Fourier transform can be expressed using Fourier components of $\zeta(t)$ as
\begin{align}
&n [\omega] = \int dt e^{i \omega t} n(t)
    \nonumber \\
= & |c_0|^2 +\int d\omega' \zeta^* [\omega - \omega']\zeta[ \omega']+c^*_0 \zeta[ \omega] +c_0 \zeta^*[ \omega] .
\end{align}

Since the Fourier transform $\zeta [\omega]$ of a Gaussian variable must also be Gaussian, polyspectra of $n(t)$ can be calculated using the expression above by applying Wick's theorem. Noting that all the anomalous correlators vanish, the only contractions that contribute would be given by terms of the following form
\begin{equation}
\overline{\zeta^*[\omega]\zeta[\omega'] }
    =   \frac{\gamma {\bar n}_\mathrm{eff}}{\left(\omega - \delta\right)^2+\left(\frac{\gamma}{2}\right)^2}   \delta (\omega +\omega')  ,
\end{equation}
which is always non-negative. It is then straightforward to show that both drive-independent and drive-dependent contributions to polyspectra must also be non-negative for all frequencies. In particular, the frequency dependence ${\tilde S}_{ \mathrm{cl}}[ {\omega}_{1},{\omega}_{2}]$ of the drive-dependent bispectrum in the classical limit 
(see main text for definition) is real and positive semidefinite, which can be explicitly written as
\begin{align}
  &{\tilde S}_{ \mathrm{cl}} 
    [ {\omega}_{1},{\omega}_{2}] \nonumber \\
=    & 
   \frac{1}{  \gamma ^2 }  \sum\limits_{\substack{\alpha  \ne \beta \\ \alpha ,\beta  = 1,2,3}  } 
    \frac{1}
    {\left[
    1+4{\left(\frac{ \omega _\alpha  +\delta}{\gamma} \right)}^2
    \right]\left[
    1+4{\left(\frac{ \omega _\beta - \delta }{\gamma} \right)}^2
    \right]}
     .
\label{AppEq:QBSdrcl}
\end{align}

\section{Temporal skewness for squeezed bath photon fluctuations}
\label{AppSec:sqbathfl}

In the main text, we show a violation of higher-order Onsager symmetry relations solely due to quantum corrections in the temporal third cumulant (skewness), which can be probed by an imaginary part in the QBS. Here we provide an example where the temporal skewness exhibits time asymmetry in both the classical and the quantum limits, and the skewness function also reveals insights into non-equilibrium dynamics in well-defined classical systems. We again consider photon shot noise in a dissipative bosonic mode, but now driven by squeezed noise. The master equation is 
\begin{equation}
\dot {\hat \rho}  =  - i [ \hat H_0 + \hat H_{\mathrm{int} },\hat \rho ] + \gamma ( {\bar n}_\mathrm{cl} + 1 ) \mathcal{D} [ \hat s_r ]\hat \rho  + \gamma {\bar n}_\mathrm{cl} \mathcal{D} [\hat s_r^\dag ]\hat \rho ,
\label{AppEq:meqSQ}
\end{equation}
where $\hat s_r = \hat c \cosh{r} + \hat c^\dag \sinh r$ denotes the squeezed bath operator. In the rotating frame, the oscillator Hamiltonian is $\hat H_0 =- \delta {\hat c^\dag }\hat c$, and its interaction with the qubit is $\hat H_{\mathrm{int}} (t) = \frac{1}{2} F(t) \hat{n}(t)  \hat \sigma_z$. Such noise model has a well-defined classical limit if we let ${\bar n}_\mathrm{cl} \to \infty$, where the bosonic mode can be equivalently described by a classical stochastic variable $c(t)$. We note that the steady state of the corresponding classical model is not thermal equilibrium, enabling a violation of Onsager-like relations even in the classical limit.

For concreteness, we again consider the temporal third cumulant $C ^{(3)}( t,t)$, which can be written as a sum of classical and quantum contributions as
\begin{equation}
C ^{(3)}( t,t)= (2{\bar n}_\mathrm{cl} + 1)^3 f( t )  \left[ {\tilde C}_{\mathrm{cl}}^{(3)}(t) - \frac{1}{(2{\bar n}_\mathrm{cl} + 1)^2}\right],
\end{equation}
where $f( t )=e^{ - \gamma | t |}\cosh (2r )/4$ is an even function of time $t$ and independent of ${\bar n}_\mathrm{cl}$. The coefficient function ${\tilde C}_{\mathrm{cl}}^{(3)}(t)$ for the classical contribution is given by
\begin{equation}
{\tilde C}_{\mathrm{cl}}^{(3)}(t) =\cosh^2 (2r)+  \frac{  \gamma ^2 \sinh^2 (2r) }{ \gamma ^2+ 4\delta ^2} [ 1+  2 \cos  ( \delta t +\delta  | t  | )   ].
\end{equation}
The situation is now reversed: the quantum correction is symmetric under time reversal $t \to -t$, whereas the classical contribution is asymmetric for a generic nonzero detuning $\delta \ne 0$.

The time asymmetry in $C ^{(3)}( t,t)$ has its roots in classical non-equilibrium dynamics: in the classical limit ${\bar n}_\mathrm{cl}  \gg 1$, we can introduce two real quadratures $x$ and $p$ defined by $c=( x + i p)/ \sqrt 2$ to describe the corresponding classical oscillator. Their dynamics satisfies the stochastic differential equations
\begin{subequations}
\begin{align}
& dx = (-\delta p - \frac{\gamma }{2} x ) dt + e^{r} \sqrt {\gamma {\bar n}_\mathrm{eff}} dW_1  , \\
& dp = ( \delta x - \frac{\gamma }{2} p ) dt + e^{-r} \sqrt {\gamma {\bar n}_\mathrm{eff}} dW_2 ,
\end{align}
\end{subequations}
where ${\bar n}_\mathrm{eff} ={\bar n}_\mathrm{cl} +1/2 \simeq {\bar n}_\mathrm{cl}$, and $dW_1 $ and $dW_2$ are independent Wiener increments. These equations formally also describe time evolution of a resonantly coupled pair of real harmonic modes, where the interaction strength is given by $|\delta |$, and each oscillator is also coupled to a thermal reservoir with thermal excitations $e^{\pm 2 r} {\bar n}_\mathrm{eff}$. This coupled two-mode system for $r \ne 0$ is a typical example of non-equilibrium system that violates detailed balance, manifested as time asymmetry in cross correlation functions $ \overline{A(0)B(t)} $~\cite{Tomita1973,Tomita1974,Carmichael2002}. Noting that $n(t)$ corresponds to the total energy in the classical limit, the skewness $C ^{(3)}( t,t)$ can then be viewed as a correlation function between energy fluctuations $\delta n(0)$ and its higher order fluctuations $[\delta  n(t )]^2$ at a different time. Thus, the time asymmetry in $C ^{(3)}( t,t)$ is again a signature of detailed balance violation, which in turn is due to the imbalanced thermal baths set by the nonzero $r$.

%%%%%%%%%%%%%%%%%%%%%%%%%%%%%%%%%%%%%%%%%%%%%%%%%%%%%%%%%%%%%%%%%%%%%%%%%%%%%%%%%%%%%%%%%%%%%%%%%%
%%%%%%%%%%%%%%%%%%%%%%%%%%%%%%%%%%%%%%%%%%%%%%%%%%%%%%%%%%%%%%%%%%%%%%%%%%%%%%%%%%%%%%%%%%%%%%%%%%
%%%%%%%%%%%%%%%%%%%%%%%%%%%%%%%%%%%%%%%%%%%%%%%%%%%%%%%%%%%%%%%%%%%%%%%%%%%%%%%%%%%%%%%%%%%%%%%%%%

%%%%%%%%%%%%%%%%%%%%%%%%%%%%%%%%
\bibliographystyle{apsrev4-1}
\pagestyle{plain}
\bibliography{refFCS}
%%%%%%%%%%%%%%%%%%%%%%%%%%%%%%%%

\end{document}